# Spectrally enhancing near-field radiative heat transfer by exciting magnetic polariton in SiC gratings

Yue Yang and Liping Wang*

*School for Engineering of Matter, Transport, and Energy*
*Arizona State University, Tempe, AZ, USA 85287*

* Corresponding author: *liping.wang@asu.edu*

**Abstract**

In the present work, we theoretically demonstrate, for the first time, that near-field radiative transport between 1D periodic grating microstructures separated by sub-wavelength vacuum gaps can be significantly enhanced by exciting magnetic resonance or polariton. Fluctuational electrodynamics that incorporates scattering matrix theory with rigorous coupled-wave analysis is employed to exactly calculate the near-field radiative heat flux between two SiC gratings. Besides the well-known coupled surface phonon polaritons (SPhP), an additional spectral radiative heat flux peak, which is due to magnetic polariton, is found within the phonon absorption band of SiC. The mechanisms, behaviors and interplays between magnetic polariton, coupled SPhP, single-interface SPhP, and Wood's anomaly in the near-field radiative transport are elucidated in detail. The findings will open up a new way to control near-field radiative heat transfer by magnetic resonance with micro/nanostructured metamaterials.

It has been demonstrated during the last decade that, radiative heat transfer could be significantly enhanced when distance between two objects is smaller than the characteristic thermal wavelength due to photon tunneling or coupling of evanescent waves. [1-3] In particular, near-field radiative heat flux could far exceed the blackbody limit by the resonant coupling of surface plasmon/phonon polaritons (SPP/SPhP) across the vacuum gap both theoretically and experimentally.[4-7] Recently, excitations of magnetic SPhP,[8,9] hyperbolic behaviors,[10-12] and epsilon-near-pole or epsilon-near-zero modes [13,14] with different types of metamaterials have also been studied to further improve the near-field heat flux. Near-field thermal radiation could find many important applications in energy-harvesting,[1,15] near-field imaging,[16] thermal modulation,[17] and thermal switch[18] and rectification.[19-21] However, other physical mechanisms have yet to be explored for enhancing near field radiative transport.

Magnetic polaritons (MP) refer to the strong coupling of external electromagnetic waves with the magnetic resonance excited inside the nanostructures. MP artificially realized with metallic micro/nanostructures has been enormously employed to control light propagation in the far-field region.[22-27] MP have been employed for designing novel metamaterials with exotic optical and radiative properties in the far field, such as selective solar absorber,[28] thermophotovoltaic emitter,[29] and switchable or tunable metamaterial.[30] Phonon-mediated MP have been excited in both SiC deep grating and binary grating configurations as well.[31] In comparison to SPP/SPhP that has been well studied for tailoring both far- and near-field thermal radiation, magnetic resonance or MP has only been investigated for controlling far-field thermal radiation while its role in near-field radiative transport has yet to be identified.

In this work, we will theoretically demonstrate, for the very first time, that near-field radiative heat transfer between two SiC grating microstructures separated by a vacuum gap *d*, as depicted in Fig. 1, could be significantly enhanced by excitation of MP. The grating period, depth, and ridge width are denoted as $\Lambda = 5$ μm, $h = 1$ μm, and $w = 4.5$ μm, respectively. The grating filling ratio is then $f = w/\Lambda = 0.9$. The temperatures of emitter and receiver are $T_1 = 400$ K and $T_2 = 300$ K, respectively. The scattering formalism [32,33] that is incorporated into fluctuational electrodynamics with rigorous coupled-wave analysis (RCWA) [34,35] is employed to exactly calculate the near-field radiative heat flux. The dielectric function of SiC is described by a Lorentz model[36] as $\varepsilon_{\text{SiC}}(\nu) = \varepsilon_\infty (1 + \frac{\nu_{\text{LO}}^2 - \nu_{\text{TO}}^2}{\nu_{\text{TO}}^2 - i\gamma\nu - \nu^2})$, where $\nu$ is the frequency in wavenumber, the high-frequency constant $\varepsilon_\infty$ is 6.7, the longitudinal optical-phonon frequency $\nu_{\text{LO}}$ is 969 cm$^{-1}$, the transverse optical-phonon frequency $\nu_{\text{TO}}$ is 793 cm$^{-1}$, and scattering rate $\gamma$ equals 4.76 cm$^{-1}$ at room temperature.

Through the exact scattering theory, near-field spectral radiative heat transfer between two gratings is expressed as [32,33,37]

$$q_\omega = \frac{1}{2\pi^3} \left[ \Theta(\omega, T_1) - \Theta(\omega, T_2) \right] \int_0^{\pi/\Lambda} \int_0^\infty \xi(\omega, k_{x0}, k_y) dk_y dk_{x0} \tag{1}$$

where $\Theta(\omega, T) = \hbar\omega / (e^{\hbar\omega/k_B T} - 1)$ is the Planck oscillator, $\omega$ is angular frequency, and $k_{x0}$ and $k_y$ are the incident wavevector components at the grating surface in x and y direction, respectively. The energy transmission coefficient $\xi(\omega, k_{x0}, k_y)$, which considers all the polarization states, is

$$\xi(\omega, k_{x0}, k_y) = \text{tr}(\mathbf{D}\mathbf{W}_1 \mathbf{D}^\dagger \mathbf{W}_2) \tag{2a}$$

$$\mathbf{D} = (\mathbf{I} - \mathbf{S}_1 \mathbf{S}_2)^{-1} \tag{2b}$$

$$\mathbf{W}_1 = \Sigma_{-1}^{pw} - \mathbf{S}_1 \Sigma_{-1}^{pw} \mathbf{S}_1^\dagger + \mathbf{S}_1 \Sigma_{-1}^{ew} - \Sigma_{-1}^{ew} \mathbf{S}_1^\dagger \tag{2c}$$

$$\mathbf{W}_2 = \Sigma_1^{pw} - \mathbf{S}_2^\dagger \Sigma_1^{pw} \mathbf{S}_2 + \mathbf{S}_2^\dagger \Sigma_1^{ew} - \Sigma_1^{ew} \mathbf{S}_2 \tag{2d}$$

where $\mathbf{S}_1 = \mathbf{R}_1$ and $\mathbf{S}_2 = e^{ik_{z0}d}\mathbf{R}_2 e^{ik_{z0}d}$. $\mathbf{R}_1$ and $\mathbf{R}_2$ are the reflection operators of the two gratings, which can be obtained through RCWA method.[34,35] The operators $\Sigma_n^{pw/ew} = \frac{1}{2}k_z^n \Pi^{pw/ew}$, where $\Pi^{pw/ew}$ are the projectors on the propagative and evanescent sectors, were clearly defined in Ref. [33]. Only 1D grating structure with periodicity along $x$ axis is considered here. According to the Bloch wave conditions, the in-plane wavevector on the interface between grating and vacuum can be expressed as

$$k_{\|,j} = \sqrt{(k_{x0} + \frac{2\pi}{\Lambda}j)^2 + k_y^2} \tag{3}$$

where $j$ (= 0, ±1, ±2, ···) refers to different diffraction order. Note that the wavevector in $x$ direction has been extended from the first Brillouin zone of $k_{x0}$ to infinity through Bloch wave conditions.

To ensure the numerical accuracy of the calculation with reasonable computational time, a total of 51 angular frequency values evenly spanned from $1.4 \times 10^{14}$ rad/s to $1.9 \times 10^{14}$ rad/s within the phonon absorption band of SiC was considered, while 31 data points were used for $k_{x0}$ and $k_y$ for calculating the spectral heat flux at each frequency after double integrations. The upper limit of $k_y$ was set as $10\omega/c$ at first with 101 data points, and then it will reduce to a cutoff value, above which the transmission coefficient is smaller than 0.5% of the maximum value when $k_{x0} = 0$. For a given frequency, the maximum diffraction order was iterated by adding 5 orders each time until the relative error between consecutive calculations is smaller than 1%. It turned out that a total of 81 diffraction orders were required to ensure the numerical convergence.

Figure 2 shows the spectral radiative heat flux between SiC gratings at several different vacuum gap distances. For $d = 1$ μm, two major peaks appear at angular frequencies of $1.61 \times 10^{14}$ rad/s (denoted as $\omega_{MP1}$) and $1.79 \times 10^{14}$ rad/s (as $\omega_{SPhP}$) in addition to a minor one at $1.75 \times 10^{14}$ rad/s (as $\omega_{MP2}$). The spectral peak at $\omega_{SPhP} = 1.79\times10^{14}$ rad/s is consistent with that between two SiC plates, which has been well known due to the coupling of SPhP between SiC across the vacuum gap.[1] However, different from the case of two SiC plates, the other broadband peak around $\omega_{MP1} = 1.61\times10^{14}$ rad/s and a small peak around $\omega_{MP2} = 1.75\times10^{14}$ rad/s only occur between two SiC gratings, which dramatically further enhance the near-field heat transfer but have not been observed before. These two additional peaks are explained as excitation of MPs in the near-field. Due to the heat transfer enhancement from excitation of SPhP and MP in the SiC gratings, the spectral heat flux between them at vacuum gap of 1 μm is about 1 order of magnitude exceeding far-field blackbody limit.

The spectral heat flux between two SiC gratings at $d = 10$ μm is also plotted in Fig. 2, which can be considered to approach the far-field radiative heat transfer between two SiC gratings due to the comparable vacuum gap distance with characteristic wavelength. There are two peaks for the heat transfer between two SiC gratings at $d = 10$ μm around $\omega_{MP1} = 1.61\times10^{14}$ rad/s and $\omega_{MP2} = 1.75\times10^{14}$ rad/s, which are consistent with the MP resonance frequencies at the far-field spectral emittance peaks of the same SiC grating studied in Ref. [31]. By comparison between the spectral heat fluxes at $d = 1$ μm and 10 μm, it can be observed that the two peaks at $\omega_{MP1} = 1.61\times10^{14}$ rad/s and $\omega_{MP2} = 1.75\times10^{14}$ rad/s occur at the same resonance frequencies in both far and near fields, suggesting that the spectral enhancement in near-field radiative transfer at $d = 1$ μm beyond the blackbody limit is associated with the excitation of MP. The near-field radiative heat transfer between SiC gratings at these resonance frequencies is almost two orders of magnitude higher than that in far field.

In order to elucidate the underlying mechanism of MP resonance in enhancing the near-field spectral heat flux, the contour plot of energy transmission coefficient at $d = 1$ μm for SiC gratings is presented in Fig. 3(a). Note that $k_y = 0$ is specified and $k_{x0}$ is normalized to the light line with $\omega_0 = 1.5\times10^{14}$ rad/s. The brightness in the figure indicates the strength of energy transmission across the vacuum gap in the near field. It can be observed that, there exist multiple enhancement bands, and when they intersect with each other, it will cause a further improvement of transmission coefficient. Note that there is a strong horizontal enhancement band at $\omega_{MP1} = 1.61\times10^{14}$ rad/s, which is independent of wavevector $k_{x0}$ and illustrated as MP resonance in near field. The independence of MP resonance location on wavevector $k_{x0}$ is the characteristic behavior of MP resonance, and has been well investigated in far field. [27,28,31]

The MP resonance frequency can also be predicted using LC circuit model by zeroing the total impedance [31]

$$Z_{tot} = i\omega(L_m + L_k - \frac{1}{\omega^2 C}) = 0 \qquad (5)$$

where $L_m$, $L_k$, $C$ are respectively mutual inductor, kinetic inductor, and capacitor, whose expressions can be found in Ref. [31]. For comparison, the resonance angular frequency is predicted to be $\omega_{LC} = 1.617\times10^{14}$ rad/s by the LC circuit model (shown as green triangles), which match very well with the horizontal enhancement band of transmission coefficient around $\omega_{MP1}$ from the exact solution, clearly confirming the excitation of MP and its enhancement of heat transfer in the near field.

On the other hand, besides the MP resonance band, there exist several other enhancement modes in Fig. 3(a) for near-field radiative transfer between SiC gratings, which are namely associated with coupled-SPhP, single-interface SPhP, and Wood's Anomaly (WA), to be discussed in detail below.

It is well known that there exists strong SPhP coupling between SiC plates in the near-field.[1] Figure 3(b) shows the contour plot of transmission coefficient between two SiC plates at $d$

= 1 μm, in which the coupled SPhP mode is folded into the first Brillouin zone according to the Bloch equation. The dispersion relation of coupled SPhP between two SiC plates is also plotted from the analytical conditions

$$\text{Symmetric mode: } \frac{k_{1z}}{\varepsilon_1} + \frac{k_{0z}}{\varepsilon_0}\coth\left(\frac{k_{0z}d}{2i}\right) = 0 \quad (6)$$

where the subscript "1" and "0" refers to SiC and vacuum, respectively. $k_z$ is the wavevector component vertical to the interface. The good agreement of the coupled-SPhP dispersion curve and transmission coefficient enhancement band confirms the SPhP coupling effect between SiC plates. However, different from the case of SiC plates, the resonance mode due to coupled SPhP between two SiC gratings, which corresponds to the enhancement band below MP in Fig. 3(a), is suppressed by MP resonance and saturates towards MP frequencies around $\omega_{MP1} = 1.61 \times 10^{14}$ rad/s.

Besides, single-interface SPhP excitation, which is not observed for plates in Fig. 3(b), is excited for near-field radiative transfer between SiC gratings in Fig. 3(a). The single SPhP can be excited at the ridge side surfaces. The dispersion relation for single-interface SPhP between SiC and vacuum is

$$k_{sp} = \frac{\omega}{c_0}\sqrt{\frac{\varepsilon_{SiC}}{\varepsilon_{SiC}+1}} \quad (7)$$

where $c_0$ is the speed of light in vacuum. The single-interface SPhP can be excited when the equation $k_{\parallel,j} = k_{sp}$ is satisfied. The dispersion relation of single SPhP at SiC plate-vacuum interface when excited is also plotted in Fig. 3(a) for the specific case of $k_y = 0$. As can be observed, the transmission coefficient enhancement due to single-interface SPhP is split into two branches. The branch at high angular frequency above MP resonance can match well with the single SPhP dispersion relation. However, similar to coupled SPhP, the branch below MP resonance seems to be suppressed by MP resonance and shifts away from dispersion relation.

The phenomenon of Wood's Anomaly (WA) is also observed in Fig. 3(a) for near-field radiative transfer between SiC gratings. Lights can be diffracted into many orders upon the grating. When the diffracted light is exactly on the grating plane, it will cause additional absorption. WA effect has been well studied for far-field optical phenomena with gratings [36], but its effect on near-field radiation has not been studied yet. The WA resonance condition can be expressed as

$$k_{\|,j} = \frac{\omega}{c_0} \tag{8}$$

The WA dispersion relation is also shown in Fig. 3(a) for the specific case of $k_y = 0$. As can be clearly observed, when WA dispersion relation intersects with other resonance modes, it will generate a brighter spot, which indicates a higher enhancement of heat transfer when WA occurs.

Each resonance mode has been discussed in detail above, and the energy transmission coefficient enhancement due to each of them has also been elucidated. Furthermore, when they couple with each other, it will cause an even much higher enhancement. These can be clearly observed from the bright hot spots of energy transmission coefficient in Fig. 3(a) around $\omega = 1.54 \times 10^{14}$ rad/s and $\omega_{MP1} = 1.61 \times 10^{14}$ rad/s. At $\omega = 1.54 \times 10^{14}$ rad/s, coupled SPhP and single SPhP intersect, while at $\omega_{MP1} = 1.61 \times 10^{14}$ rad/s, all the three modes of MP, single SPhP and WA couple with each other. The magnitude of heat transfer between two SiC gratings can be 1 order larger than that between two SiC plates. Note that the results shown in Fig. 3 are all based on the assumption of $k_y = 0$. Although it is not presented in the paper, the case of $k_y \neq 0$ has also been investigated. The interactions between different physical resonance modes are very similar to the case of $k_y = 0$, but the folding positions of WA, single-interface SPhP and coupled SPhP will shift to higher frequencies due to $k_y \neq 0$.

In order to further understand the behaviors of different resonance modes in near-field radiative transport between grating structures, the geometric effect is studied. Figure 4(a) shows the contour plot of the transmission coefficient at $d = 1$ μm and $k_y = 0$ when the grating period is decreased from $\Lambda = 5$ μm to 3 μm in comparison to Fig. 3(a), while other geometric parameters like $f = 0.9$ and $h = 1$ μm are kept the same. The flat MP resonance mode shifts a slightly lower frequency around $\omega_{MP1} = 1.58 \times 10^{14}$ rad/s. As studied from the far-field radiative properties, the grating period has little effect in influencing the MP resonance condition in SiC gratings. However, as the filling ratio is fixed, smaller period would result in the decrease of groove width, which shifts the MP resonance frequency to lower values.[31] The LC model also predicts a smaller MP resonance frequency of $\omega_{LC} = 1.565 \times 10^{14}$ rad/s with $\Lambda = 3$ μm.

On the other hand, the resonance conditions of coupled SPhP, single-interface SPhP and WA strongly depend on the grating period, as inferred from Eqs. (6-8). The edge of the first Brillouin zone (i.e. $k = \pi/\Lambda$) becomes larger when $\Lambda$ decreases from 5 μm to 3 μm. Due to the intersection with the MP resonance mode, the coupled SPhP mode is split into one high- and the other low-frequency branches around $\omega_{MP1}$. The enhancement of transmission coefficient due to the single-interface SPhP mode can be observed, which matches well with its dispersion curve. It can be clearly seen that, strong enhancement in near-field radiative transfer occurs when MP or coupled SPhP inter-couples with single-interface SPhP and WA.

When the grating height is reduced from $h = 1$ μm to 0.5 μm, the MP1 resonance band shifts to higher frequencies around $\omega_{MP1} = 1.66 \times 10^{14}$ rad/s as seen from the transmission coefficient in Fig. 4(b). Note that, the MP resonance condition has strong dependence on $h$ as both demonstrated by RCWA and LC model in the far field. On the other hand, the single-interface SPhP and WA resonance modes are little affected by the grating depth with excellent agreement with their respective dispersion relations, from which $h$ is not involved nevertheless. Similar depth-independent behavior should exist for coupled SPhP as well. However, due to its strong inter-coupling with the MP mode, the low-frequency branch shifts to higher frequency as MP resonance frequency increases, while the high-frequency coupled SPhP mode merges with the MP mode. A broader spectral enhancement on the near-field radiative transfer occurs when the MP, coupled SPhP, single-interface SPhP and WA modes intersect.

In summary, we have theoretically demonstrated that, near-field radiative heat flux could be greatly enhanced by exciting MP or magnetic resonance in SiC gratings. The rigorous calculation has clearly shown that, multiple resonance modes including MP, single-interface SPhP, and WA in addition to the well-known coupled SPhP exist and enhance the radiative transport between microstructured grating structures separated by subwavelength vacuum gaps. The underlying mechanisms of MP and its interplay with other resonance modes in the near field have been confirmed and elucidated with the fluctuational electrodynamics, analytical *LC* model, and dispersion relations. The understanding gained here will undoubtedly lead to the spectral control of near-field thermal radiation with magnetic polaritons for energy-harvesting applications like thermophotovoltaic energy conversion and radiation-based thermal management.

ACKNOWLEDGMENT

YY and LW are grateful to the support from the ASU New Faculty Startup fund. YY would like to thank the partial support from the University Graduate Fellowship offered by the ASU Fulton Schools of Engineering.

REFERENCES

1. S. Basu, Z. M. Zhang, and C. J. Fu, Int. J. Energy Res. **33**, 1203 (2009).
2. D. G. Cahill, P. V. Braun, G. Chen, D. R. Clarke, S. H. Fan, K. E. Goodson, P. Keblinski, W. P. King, G. D. Mahan, and A. Majumdar, Appl. Phys. Rev. **1**, 011305 (2014).
3. D. G. Cahill, W. K. Ford, K. E. Goodson, G. D. Mahan, A. Majumdar, H. J. Maris, R. Merlin, and S. R. Phillpot, J. Appl. Phys. **93**, 793 (2003).
4. C. Fu and Z. Zhang, Int. J. heat Mass Tran. **49**, 1703 (2006).
5. S. Shen, A. Mavrokefalos, P. Sambegoro, and G. Chen, Appl. Phys. Lett. **100**, 233114 (2012).
6. S. Shen, A. Narayanaswamy, and G. Chen, Nano Lett. **9**, 2909 (2009).
7. Y. Yang, S. Basu, and L. Wang, Appl. Phys. Lett. **103**, 163101 (2013).
8. S. Basu and M. Francoeur, Appl. Phys. Lett. **99**, 143107 (2011).
9. S. J. Petersen, S. Basu, and M. Francoeur, Photonics Nanostruct. **11**, 167 (2013).
10. S.-A. Biehs, M. Tschikin, and P. Ben-Abdallah, Phys. Rev. Lett. **109**, 104301 (2012).
11. C. Cortes, W. Newman, S. Molesky, and Z. Jacob, J. Opt. **14**, 063001 (2012).
12. Y. Guo, C. L. Cortes, S. Molesky, and Z. Jacob, Appl. Phys. Lett. **101**, 131106 (2012).
13. S. Vassant, J.-P. Hugonin, F. Marquier, and J.-J. Greffet, Opt. Express **20**, 23971 (2012).
14. S. Molesky, C. J. Dewalt, and Z. Jacob, Opt. Express **21**, A96 (2013).
15. K. Park, S. Basu, W. P. King, and Z. M. Zhang, J. Quant. Spectrosc. RA. **109**, 305 (2008).
16. K. Hoshino, A. Gopal, M. S. Glaz, D. A. Vanden Bout, and X. Zhang, Appl. Phys. Lett. **101**, 043118 (2012).
17. P. Ben-Abdallah and S.-A. Biehs, Phys. Rev. Lett. **112**, 044301 (2014).
18. Y. Yang, S. Basu, and L. Wang, J. Quant. Spectrosc. RA. **103**, 163101 (2014).
19. C. R. Otey, W. T. Lau, and S. Fan, Phys. Rev. Lett. **104**, 154301 (2010).
20. L. P. Wang and Z. M. Zhang, Nanosc. Microsc. Therm. **17**, 337 (2013).
21. Y. Yang, S. Basu, and L. Wang, Appl. Phys. Lett. **103**, 163101 (2013).
22. J. B. Pendry, A. J. Holden, D. Robbins, and W. Stewart, Microwave Theory and Techniques, IEEE Trans. **47**, 2075 (1999).
23. J. B. Pendry, Phys. Rev. Lett. **85**, 3966 (2000).
24. V. M. Shalaev, W. Cai, U. K. Chettiar, H.-K. Yuan, A. K. Sarychev, V. P. Drachev, and A. V. Kildishev, Opt. Lett. **30**, 3356 (2005).
25. V. M. Shalaev, Nat. Photonics **1**, 41 (2007).
26. B. Lee, L. Wang, and Z. Zhang, Opt. Express **16**, 11328 (2008).
27. L. Wang and Z. Zhang, Appl. Phys. Lett. **95**, 111904 (2009).
28. H. Wang and L. Wang, Opt. Express **21**, A1078 (2013).
29. B. Zhao, L. Wang, Y. Shuai, and Z. M. Zhang, Int. J. Heat Mass Tran. **67**, 637 (2013).
30. H. Wang, Y. Yang, and L. Wang, Appl. Phys. Lett. **105**, 071907 (2014).
31. L. P. Wang and Z. M. Zhang, Opt. Express **19**, A126 (2011).
32. A. Lambrecht and V. N. Marachevsky, Phys. Rev. Lett. **101**, 160403 (2008).
33. J. Lussange, R. Guérout, F. S. S. Rosa, J. J. Greffet, A. Lambrecht, and S. Reynaud, Phys. Rev. B **86**, 085432 (2012).
34. L. Li, J. Opt. Soc. Am. A **13**, 1870 (1996).
35. M. G. Moharam, E. B. Grann, D. A. Pommet, and T. K. Gaylord, J. Opt. Soc. Am. A **12**, 1068 (1995).
36. Z. M. Zhang, *Nano/Microscale Heat Transfer* (McGraw-Hill, New York, 2007).
37. X. Liu and Z. Zhang, Appl. Phys. Lett. **104**, 251911 (2014)

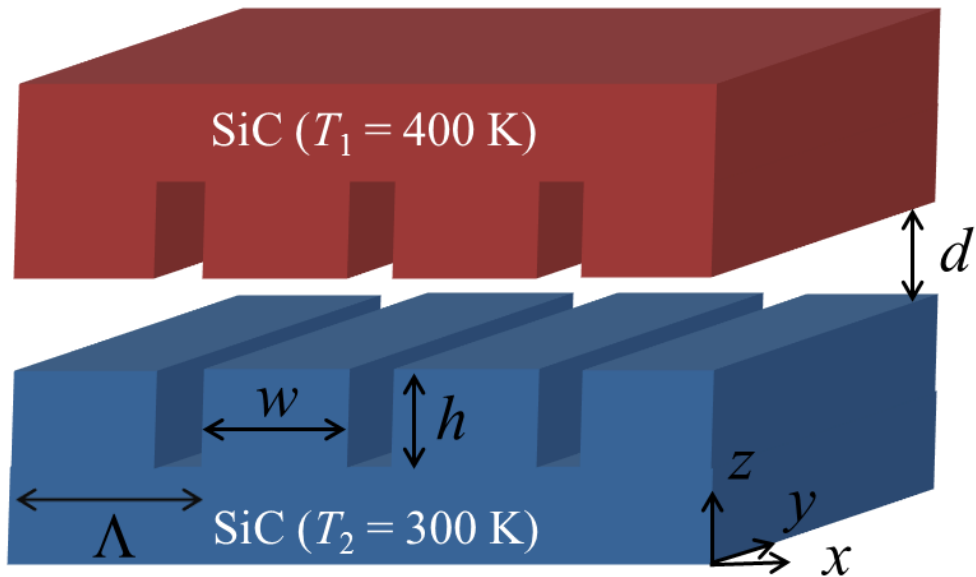

Fig. 1. Schematic of radiative heat transfer between two symmetric SiC deep gratings with the grating parameters of period ($\Lambda$), depth ($h$), and ridge width ($w$). The emitter and receiver temperatures are respectively set as $T_1$ = 400 K, and $T_2$ = 300 K in this study. The vacuum gap distance is denoted as $d$.

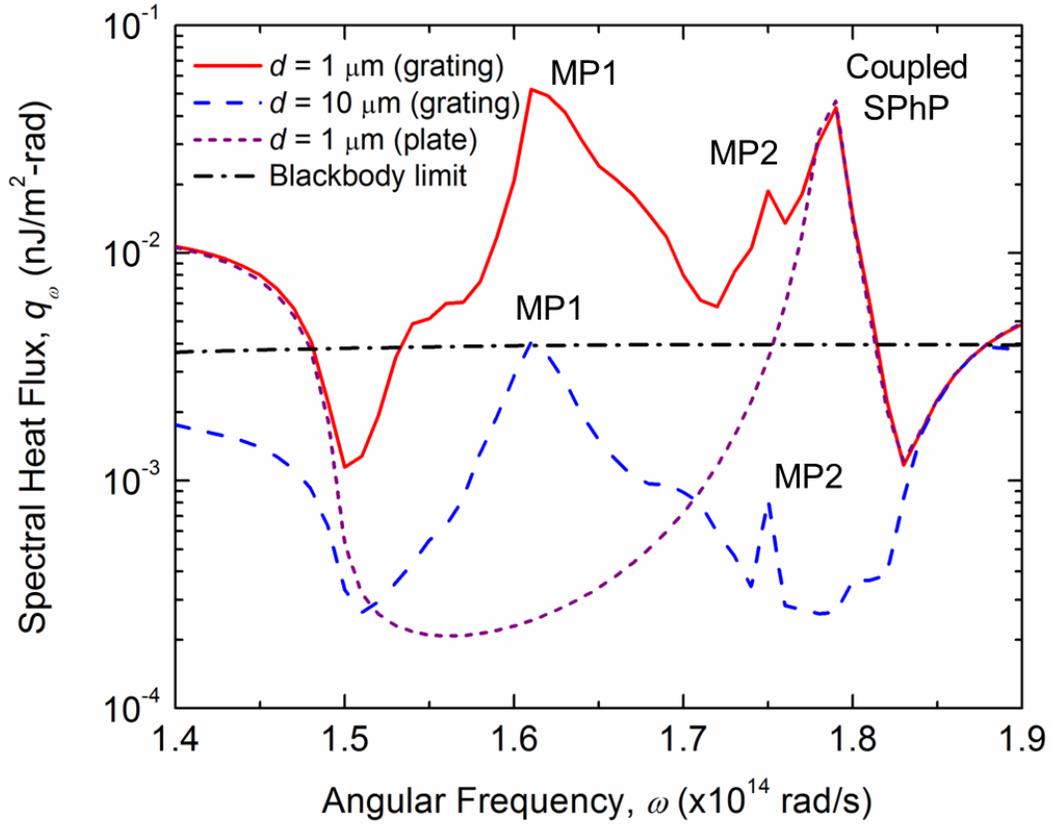

Fig. 2. Spectral heat fluxes between two SiC gratings (grating) at different vacuum gaps. The spectral heat fluxes between two SiC plates (plate) at $d = 1$ μm, and for blackbody limit are also presented. The emitter and receiver temperatures are set as 400 K and 300 K, respectively. The grating parameters are set as $\Lambda = 5$ μm, $h = 1$ μm, $w = 4.5$ μm.

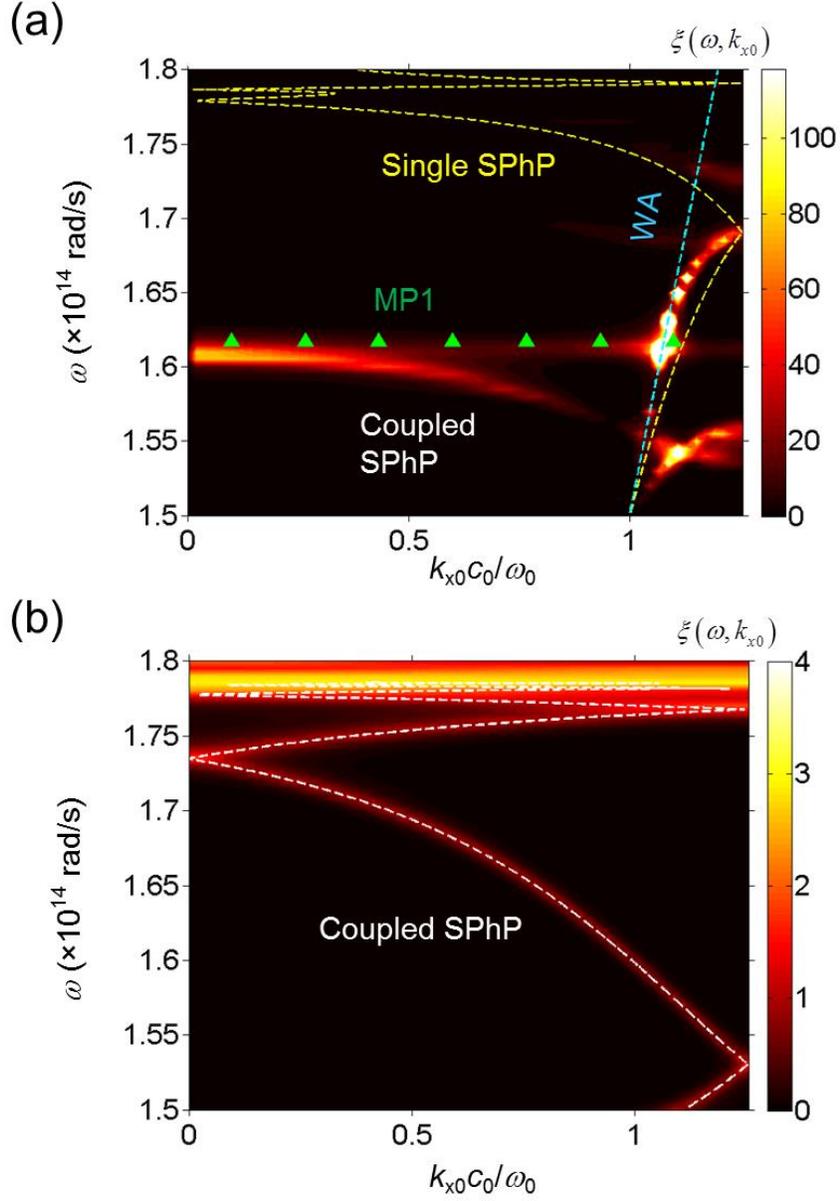

Fig. 3. Contour plot of energy transmission coefficient ($\xi$) between (a) two symmetric SiC gratings at vacuum gap of 1 μm with grating parameters setup of $\Lambda = 5$ μm, $h = 1$ μm, $w = 4.5$ μm and (b) two SiC plates at vacuum gap of 1 μm. Note that $k_y = 0$ is assumed in this plot and $k_{x0}$ is normalized ($\omega_0 = 1.5 \times 10^{14}$ rad/s). The dispersion relations for multiple physical resonance modes, such as coupled SPhP between two SiC plates at $d = 1$ μm, single SPhP at one SiC plate-vacuum interface, far-field MP of single SiC grating, and Wood's Anomaly are also plotted.

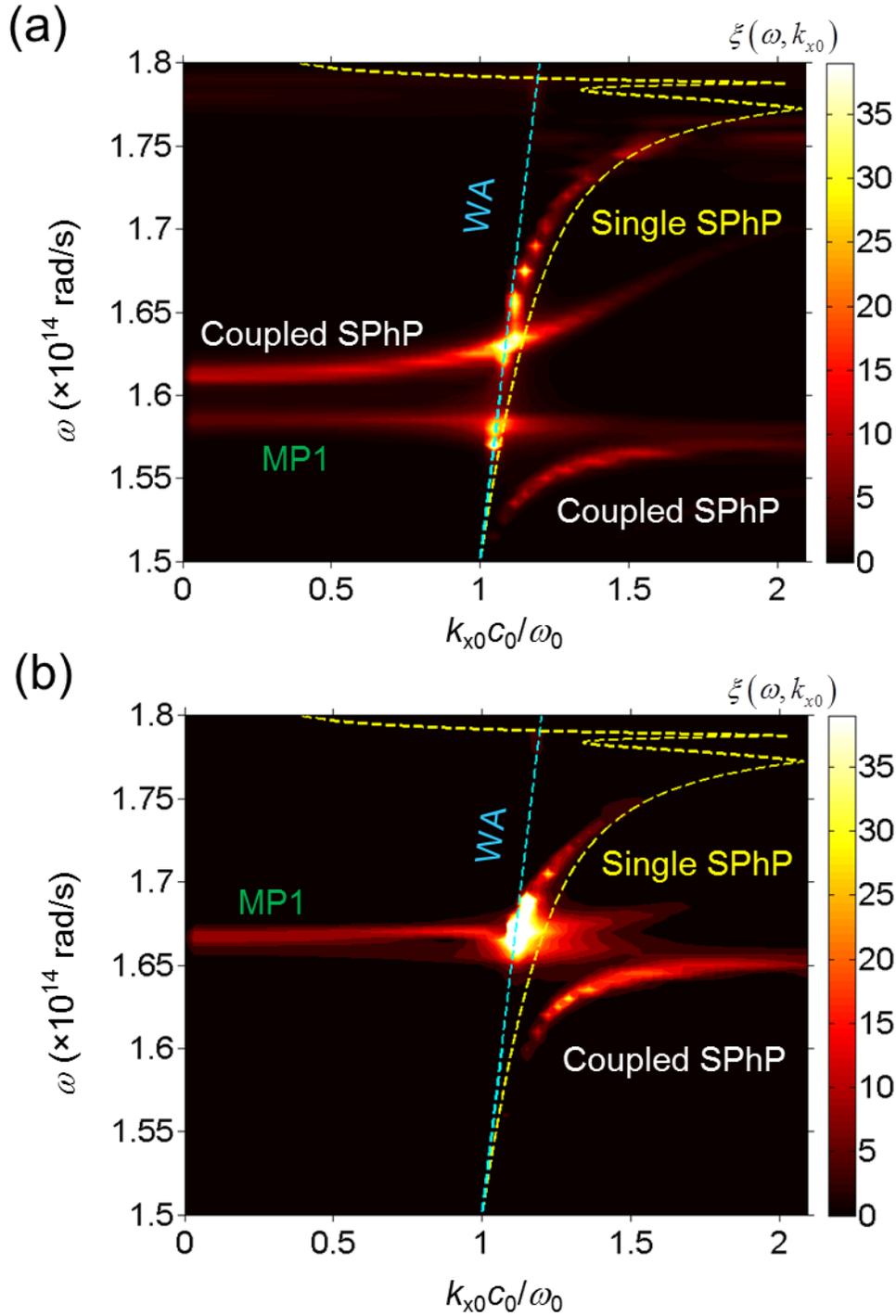

Fig. 4. Contour plot of transmission coefficient ($\xi$) between two symmetric SiC gratings with different geometries of (a) $\Lambda = 3$ μm, $h = 1$ μm; (b) $\Lambda = 3$ μm, $h = 0.5$ μm. Note that ratio of grating ridge width and period, and vacuum gap distance are assumed as constants of 0.9 and 1 μm for all cases.